\documentclass[12pt]{iopart}

\usepackage{amsfonts}
\usepackage{amssymb,graphics,epsfig,subfigure}

\begin{document}

\title[]{P-V criticality of conformal anomaly corrected AdS black holes}

\author{Jie-Xiong Mo $^{a,b}$, Wen-Biao Liu $^a$}

\address{$^a$ Department of Physics, Institute of Theoretical Physics, Beijing Normal University, Beijing, 100875, China\\
 $^b$ Institute of Theoretical Physics, Lingnan Normal University, Zhanjiang, Guangdong, 524048, China}
\ead{wbliu@bnu.edu.cn}
\vspace{10pt}

\begin{abstract}
The effects of conformal anomaly on the thermodynamics of black holes are investigated in this Letter from the perspective of $P-V$ criticality of AdS black holes. Treating the cosmological constant as thermodynamic pressure, we extend the recent research to the extended phase space. Firstly, we study the $P$-$V$ criticality of the uncharged AdS black holes with conformal anomaly and find that conformal anomaly does not influence whether there exists Van der Waals like critical behavior. Secondly, we investigate the $P$-$V$ criticality of the charged cases and find that conformal anomaly influences not only the critical physical quantities but also the ratio $\frac{P_cr_c}{T_c}$. The ratio is no longer a constant as before but a function of conformal anomaly parameter $\tilde{\alpha}$. We also show that the conformal parameter should satisfy a certain range to guarantee the existence of critical point that has physical meaning. Our results show the effects of conformal anomaly.
\end{abstract}

\section{Introduction}
     Conformal anomaly is such an
important concept that it has lots of applications in
quantum field theory, cosmology and black hole
physics. So it would be of great interest to investigate its effects on the thermodynamics of black holes. Many efforts have been devoted
to this issue. Cai et al.~\cite{Cai9999} investigated the thermodynamics of static and spherically symmetric black holes with
conformal anomaly. In addition to Bekenstein-Hawking area
entropy, an extra logarithmic correction was found. The entropy was also studied by Li who took an approach of quantum tunneling~\cite{Liran}. Considering the back reaction through the conformal anomaly, phase transitions of Schwarzschild black hole were investigated~\cite{Son9999} and an additional phase transition was discovered. Ehrenfest equations were also studied~\cite{Chenghongbo}. Moreover, we studied the phase structures of black holes with conformal anomaly~\cite{Wenbiao5}and found that they have much richer phase structures than black holes without conformal anomaly. Recently, Cai~\cite{Cai99999} obtained analytical AdS black hole solutions with conformal anomaly. Hawking-Page transition was investigated and novel features due to conformal anomaly were found in black holes with a negative constant curvature horizon. Although many efforts have been made, all of them were carried out in the non-extended phase space. In this Letter, we would like to extend the research of Ref.~\cite{Cai99999} to the extended phase space, hoping to observe more evidences of the effects of conformal anomaly on the thermodynamics of black holes.

     The extended phase space refers to the phase space which includes the pressure and volume as thermodynamic variables. Thermodynamics in the extended phase space has gained more and more attention recently~\cite{Kastor}-~\cite{Dolan97}. The main idea is to treat the cosmological constant as thermodynamic pressure and the conjugate quantity as volume. The motivation comes from the consideration of variation of cosmological constant to make the first law of black hole thermodynamics consistent with the Smarr relation. The critical behavior of black holes in the extended phase space, especially $P-V$ criticality, has been extensively investigated~\cite{Kubiznak}-~\cite{Frassino}. For nice review, see Ref.~\cite{Altamirano3,Dolan97}.

    The outline of this Letter is as follows. In Sec.\ref{Sec2}, thermodynamics of AdS black holes with conformal anomaly will be briefly studied in the extended phase space. Then $P$-$V$ criticality of the uncharged AdS black holes with conformal anomaly will be investigated in Sec.\ref{Sec3} while the charged cases will be discussed in Sec.\ref {Sec4}. To observe the effects of conformal anomaly, comparisons will be made between AdS black holes with conformal anomaly and those without it. In the end, conclusions will be drawn in Sec.\ref{Sec5}.

\section{Thermodynamics of conformal anomaly corrected AdS black hole in the extended phase space}
\label{Sec2}
The newly derived conformal anomaly corrected AdS black holes' metric can be written as~\cite{Cai99999}
\begin{equation}
ds^2=-f(r)dt^2+f(r)^{-1}dr^2+r^2d\Omega^2_{2k},\label{1}\\
\end{equation}
where
\begin{equation}
f(r)=k-\frac{r^2}{4\tilde{\alpha}}\left(1-\sqrt{1+\frac{8\tilde{\alpha}}{l^2}-\frac{16\tilde{\alpha}GM}{r^3}+\frac{8\tilde{\alpha}Q^2}{r^4}}\right).\label{2}\\
\end{equation}
$d\Omega^2_{2k}$ denotes the line element of two dimensional Einstein space with constant scalar curvature $2k$, where $k$ can be taken as $0, 1$ or $-1$ without loss of generality~\cite{Cai99999}. It was argued that the integration constant $M$ and $Q$ should be interpreted as
the mass of black holes and the U(1) conserved charge of the conformal field theory respectively~\cite{Cai99999}. When $\tilde{\alpha}\rightarrow 0$~\cite{Cai99999}, the solutions reduces to the Reissner-Nordstr\"{o}m-AdS (RN-AdS) black holes as
\begin{equation}
f(r)=k+\frac{r^2}{l^2}-\frac{2GM}{r}+\frac{Q^2}{r^2}.\label{3}\\
\end{equation}
One can obtain the event horizon radius $r_+$ by solving the equation $f(r_+) =0$ for the largest root. With the event horizon radius, one can derive the expression of the mass of black holes as
\begin{equation}
M=\frac{r_+^4+kl^2r_+^2-2k^2l^2\tilde{\alpha}+l^2Q^2}{2l^2r_+}.\label{4}\\
\end{equation}
Note that $G$ has been set to one. The Hawking temperature can be obtained as
\begin{equation}
T=\frac{f'(r_+)}{4\pi}=\frac{r_+}{4\pi(r_+^2-4k\tilde{\alpha})}\left(k+\frac{3r_+^2}{l^2}-\frac{Q^2}{r_+^2}+\frac{2k^2\tilde{\alpha}}{r_+^2}\right).\label{5}
\end{equation}
The entropy can be derived as
\begin{equation}
S=\int \frac{1}{T}\left(\frac{\partial M}{\partial r_+}\right)dr_+=\pi r_+^2-8\pi k\tilde{\alpha}\ln r_++S_0 .\label{6}
\end{equation}
Thermodynamic pressure and volume in the extended phase space can be defined as
\begin{eqnarray}
P&=&-\frac{\Lambda}{8\pi}=\frac{3}{8\pi l^2},\label{7}
\\
V&=&\left(\frac{\partial M}{\partial P}\right)_{S,Q}.\label{8}
\end{eqnarray}
Note that the second equation in Eq.(\ref{7}) holds for four dimensional black holes.
Utilizing Eqs. (\ref{4}) and (\ref{7}), one can reorganize the expression of the mass as
\begin{equation}
M=\frac{8P\pi r_+^4+3kr_+^2-6k^2\tilde{\alpha}+3Q^2}{6r_+},\label{9}
\end{equation}
from which one can easily derive the thermodynamic volume as
\begin{equation}
V=\frac{4\pi r_+^3}{3}.\label{10}
\end{equation}
Note that the entropy gains an extra logarithmic term due to the effect of conformal anomaly while the thermodynamic volume is the same as that of RN-AdS black holes.
It is quite easy to verify that the first law of black hole thermodynamics in the extended phase space and the Smarr relation can be written as
\begin{eqnarray}
dM&=&TdS+\Phi dQ+VdP,\label{11}
\\
M&=&2TS+\Phi Q-2VP,\label{12}
\end{eqnarray}%
where
\begin{equation}
\Phi=\left(\frac{\partial M}{\partial Q}\right)_{S,P}=\frac{Q}{r_+}.\label{13}
\end{equation}

\section{$P$-$V$ criticality of uncharged AdS black holes with conformal anomaly}
\label{Sec3}
In this section, we focus on the uncharged case $Q=0$.
Substituting Eq. (\ref{7}) and $Q=0$ into Eq. (\ref{5}), we obtain the Hawking temperature for the uncharge case as
\begin{equation}
T=\frac{kr_+^2+8P\pi r_+^4+2k^2\tilde{\alpha}}{4\pi r_+(r_+^2-4k\tilde{\alpha})},\label{14}
\end{equation}
from which we can derive the equation of state as
\begin{equation}
P=\frac{T}{2r_+}-\frac{k}{8\pi r_+^2}-\frac{2kT\tilde{\alpha}}{r_+^3}-\frac{2k^2\tilde{\alpha}}{8\pi r_+^4}.\label{15}
\end{equation}
The critical point can be defined as follows
\begin{eqnarray}
\left.\frac{\partial P}{\partial r_+}\right|_{r=r_c,T=T_c}&=&0,\label{16}\\
\left.\frac{\partial ^2P}{\partial r_+^2}\right|_{r=r_c,T=T_c}&=&0,\label{17}
\end{eqnarray}
where the subscript "c" denotes the physical quantities at the critical point.
Utilizing  Eqs. (\ref{15}) and (\ref{16}), one can get
\begin{equation}
T_c=\frac{kr_c^2+4k^2\tilde{\alpha}}{2\pi r_c(r_c^2-12k\tilde{\alpha})},\label{18}
\end{equation}
Solving Eq. (\ref{17}) and then substituting Eq. (\ref{18}) into the result, one can obtain
\begin{equation}
k\left(r_c^{4}+24k\tilde{\alpha}r_c^2-48k^2\tilde{\alpha}^2\right)=0.\label{19}
\end{equation}
Except $k=0$ (we would not consider this case for the Hawking temperature is zero), Eq. (\ref{19}) has two positive roots for $r_c$, namely,
\begin{equation}
r_{c}=2\sqrt{(-3k\mp 2\sqrt{3})k\tilde{\alpha}},\label{20}
\end{equation}
where "$-$" corresponds to the case $k=-1$ while "$+$" corresponds to the case $k=1$. However, substituting Eq. (\ref{20}) back into Eq. (\ref{18}), one can derive that
\begin{equation}
T_{c}=-\frac{\sqrt{(3\mp2\sqrt{3})k\tilde{\alpha}}}{12\pi \tilde{\alpha}},\label{21}
\end{equation}
where "$-$" corresponds to the case $k=-1$ while "$+$" corresponds to the case $k=1$. It is obvious that the Hawking temperature at these two critical points is negative. So these two critical points do not make any sense physically and there would be no Van der Waals like critical behavior for the uncharged case. This is in accord with Schwarzschild AdS black holes, implying that the conformal anomaly does not influence whether there exists Van der Waals like critical behavior.

\section{$P$-$V$ criticality of charged AdS black holes with conformal anomaly}
\label{Sec4}
Utilizing Eqs. (\ref{5}) and (\ref{7}), one can obtain the equation of state of charged AdS black holes with conformal anomaly as
\begin{equation}
P=\frac{T}{2r_+}-\frac{k}{8\pi r_+^2}-\frac{2kT\tilde{\alpha}}{r_+^3}+\frac{Q^2-2k^2\tilde{\alpha}}{8\pi r_+^4}.\label{22}
\end{equation}
Substituting Eq. (\ref{22}) into Eq. (\ref{16}), one can get
\begin{equation}
T_c=\frac{kr_c^2+4k^2\tilde{\alpha}-2Q^2}{2\pi r_c(r_c^2-12k\tilde{\alpha})}.\label{23}
\end{equation}
Utilizing Eqs. (\ref{17}), (\ref{22}) and (\ref{23}), one can obtain
\begin{equation}
kr_c^{4}+(24k^2\tilde{\alpha}-6Q^2)r_c^2-48k^3\tilde{\alpha}^2+24kQ^2\tilde{\alpha}=0.\label{24}
\end{equation}
Solving the above equation, one can obtain the positive roots as
\begin{equation}
r_{c}=\sqrt{\frac{3Q^2}{k}-12k\tilde{\alpha}\mp\frac{\sqrt{3}\sqrt{3Q^4-32k^2Q^2\tilde{\alpha}+64k^4\tilde{\alpha}^2}}{k}},\label{25}
\end{equation}
where "$-$" corresponds to $r_{c1}$ while "$+$" corresponds to $r_{c2}$. Note that we also omit the case $k=0$ when we solve Eq. (\ref{24}) because the Hawking temperature is negative.

\begin{figure*}
\centerline{\subfigure[]{\label{1a}
\includegraphics[width=8cm,height=6cm]{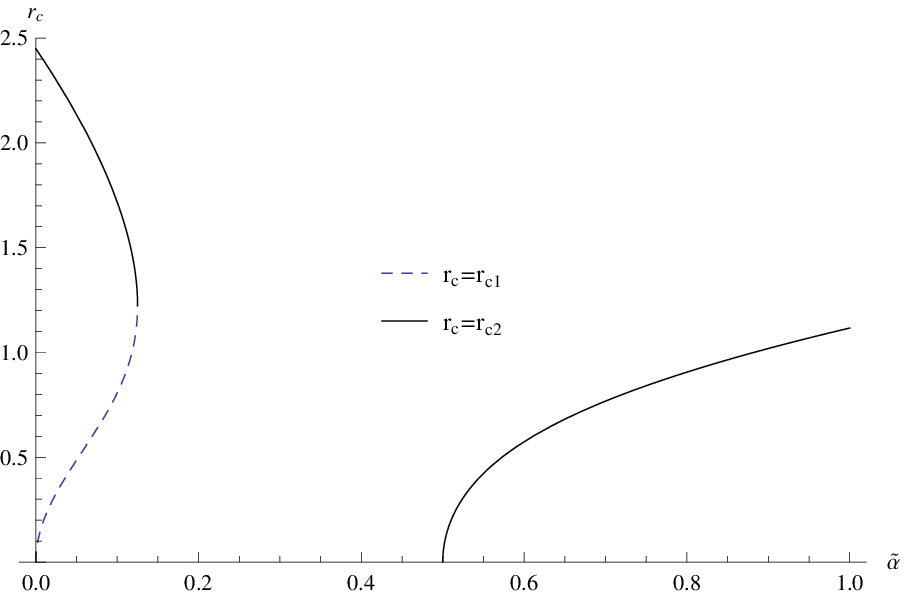}}
\subfigure[]{\label{1b}
\includegraphics[width=8cm,height=6cm]{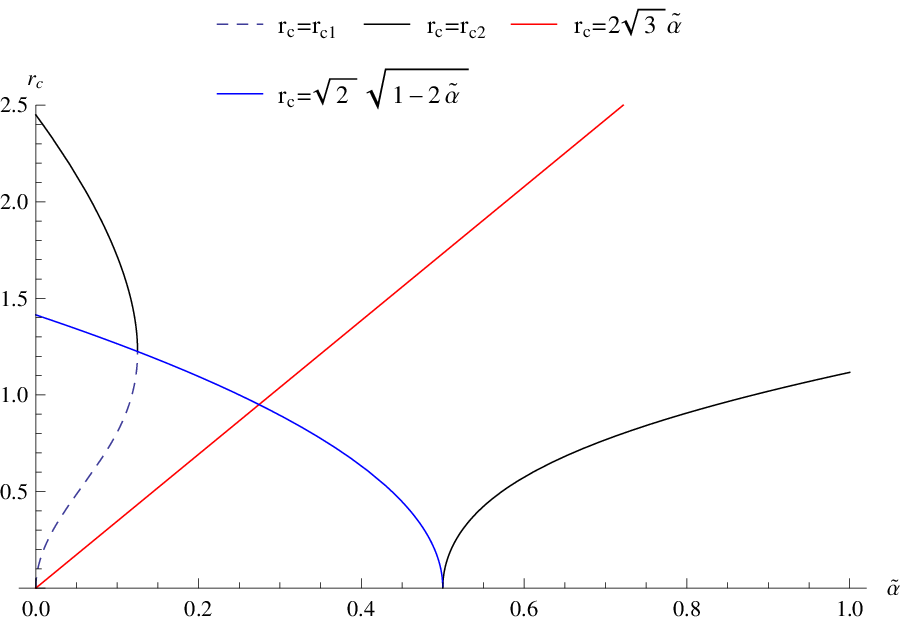}}}
 \caption{(a) $r_c$ vs. $\tilde{\alpha}$ for $k=1, Q=1\;\;\;$(b) Curve constrained by positive Hawking temperature} \label{fg1}
\end{figure*}

When $k=1$, $\tilde{\alpha}=0$, Eq. (\ref{25}) reduces to $r_c=\sqrt{6}Q$, recovering the result of RN-AdS black holes in former literature~\cite{Kubiznak}. When $\tilde{\alpha}\neq0$, one can easily draw the conclusion  from Eq. (\ref{25}) that the location of critical point relies on the conformal anomaly parameter $\tilde{\alpha}$. To observe the effect of conformal anomaly, we fix $k=1, Q=1$ and show the behavior of $r_c$ for different $\tilde{\alpha}$ in Fig.\ref{1a}. The black holes with conformal anomaly have much richer phase structure. They may have none, one or two critical points due to different $\tilde{\alpha}$. However, one should check the Hawking temperature to make sure whether the critical points are in the physical region. The positive Hawking temperature means
\begin{equation}
\left.T_c\right|_{k=1,Q=1}=\frac{r_c^2+4\tilde{\alpha}-2}{2\pi r_c(r_c^2-12\tilde{\alpha})}>0
,\label{26}
\end{equation}
which can be solved as
\begin{equation}
r_c>A \;or \;0<r_c<B,\label{27}
\end{equation}
where $A=max\{2\sqrt{3}\tilde{\alpha},\sqrt{2-4\tilde{\alpha}}\}, B=min\{2\sqrt{3}\tilde{\alpha},\sqrt{2-4\tilde{\alpha}}\}$. To witness the constraint of the Hawking temperature intuitively, we plot Fig.\ref{1b}.
From Eq. (\ref{27}) and Fig.\ref{1b}, we find that only one critical point exists under the constraining condition of Hawking temperature. And the conformal anomaly parameter $\tilde{\alpha}$ should satisfy $0<\tilde{\alpha}<\frac{1}{8}$.

\begin{figure*}
\centerline{\subfigure[]{\label{2a}
\includegraphics[width=8cm,height=6cm]{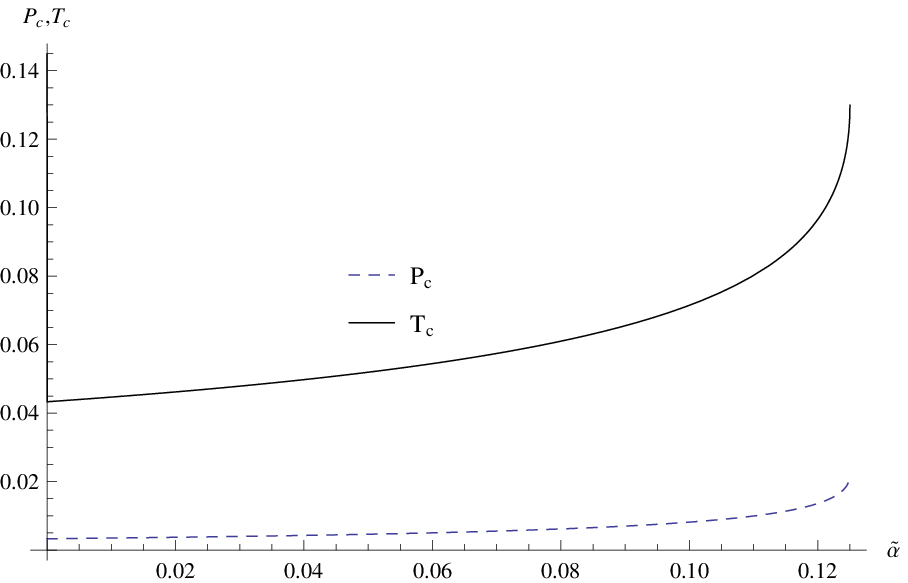}}
\subfigure[]{\label{2b}
\includegraphics[width=8cm,height=6cm]{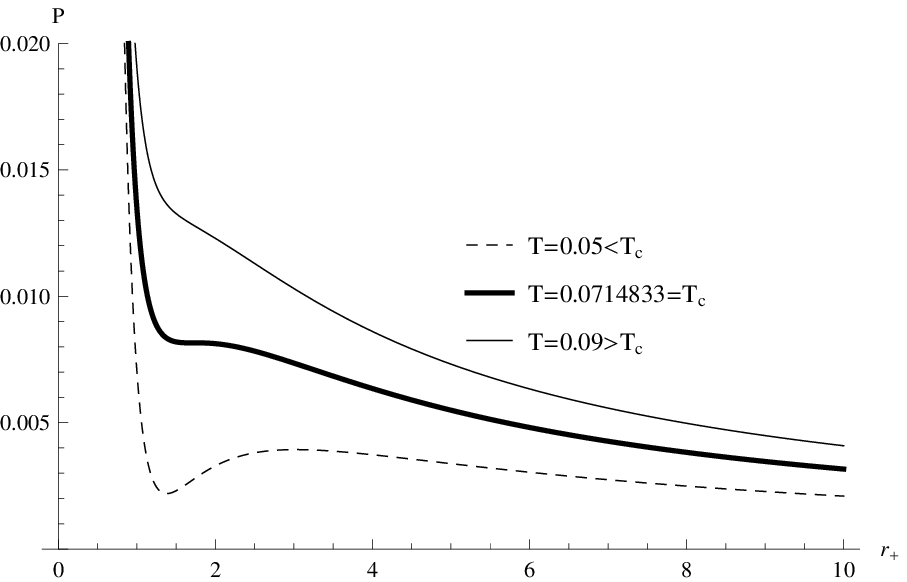}}}
 \caption{(a) $P_c, T_c$ vs. $\tilde{\alpha}$ for $k=1, Q=1\;\;\;$ (b)$P$ vs. $r_+$ for $k=1, Q=1, \tilde{\alpha}=0.1$} \label{fg2}
\end{figure*}

From Eqs. (\ref{22}), (\ref{25}), (\ref{26}) and Fig.\ref{1b}, one can derive the explicit expressions for all the critical quantities as
\begin{eqnarray}
\left.r_c\right|_{k=1,Q=1}&=&\sqrt{3-12\tilde{\alpha}+\sqrt{9-96\tilde{\alpha}+192\tilde{\alpha}^2}},
\nonumber
\\
\left.T_c\right|_{k=1,Q=1}&=&\frac{\sqrt{3-12\tilde{\alpha}+\sqrt{9-96\tilde{\alpha}+192\tilde{\alpha}^2}}}{192\pi \tilde{\alpha}^2}
\nonumber
\\
&\,&\times(3-16\tilde{\alpha}-\sqrt{9-96\tilde{\alpha}+192\tilde{\alpha}^2}),
\nonumber
\\
\left.P_c\right|_{k=1,Q=1}&=&\frac{6-18\tilde{\alpha}+\sqrt{9-96\tilde{\alpha}+192\tilde{\alpha}^2}}{24\pi \left(3-12\tilde{\alpha}+\sqrt{9-96\tilde{\alpha}+192\tilde{\alpha}^2}\right)^2}.
\nonumber
\\
\label{28}
\end{eqnarray}
The behavior of $P_c$ and $T_c$ is depicted in Fig.\ref{2a}. From Figs.\ref{1b} and \ref{2a}, we can see clearly that with the increasing of $\tilde{\alpha}$, both $T_c$ and $P_c$ increase while $r_c$ decreases. $P-r_+$ diagram for a specific case $\tilde{\alpha}=0.1$ is shown in Fig.\ref{2b}. When the temperature is lower than the critical temperature, the isotherm not only has stable small radius branch and stable large radius branch but also has the unstable medium radius branch.  However, when the temperature is higher than the critical temperature, the above phenomenon disappears.

We also plot $P-T$ curve in Fig.\ref{3} for three different choices of $\tilde{\alpha}$. Namely, $\tilde{\alpha}=0$, $\tilde{\alpha}=0.05$, $\tilde{\alpha}=0.1$. Similar curves as the $P-T$ curve of liquid-gas
phase transition can be observed. When $\tilde{\alpha}=0$, the curve recovers that of RN-AdS black hole shown in former literature~\cite{Kubiznak}. The effect of conformal anomaly is reflected in different endpoints of the curves. These endpoints correspond to the critical points. Since we have demonstrate analytically that physical quantities vary with the conformal anomaly parameter, the graphical results match the analytical results quite well.

\begin{figure*}
\centerline{\subfigure[]{\label{3a}
\includegraphics[width=8cm,height=6cm]{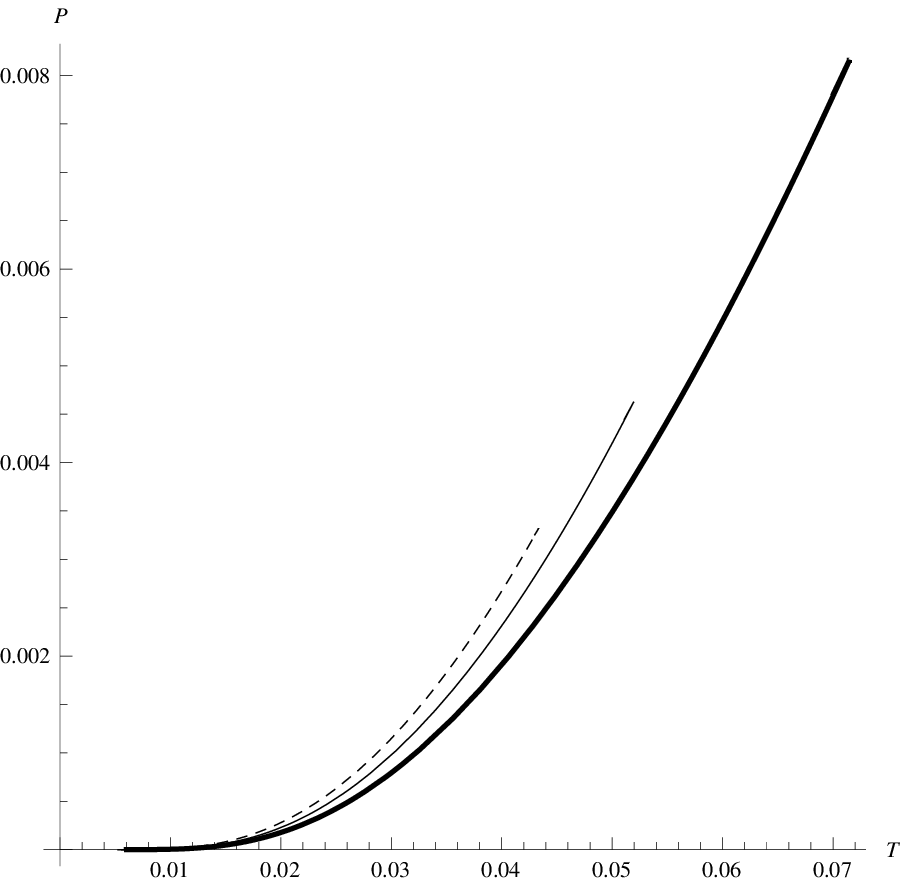}}}
 \caption{$P$ vs. $T$ when $k=1, Q=1$. Dashed curve for $\tilde{\alpha}=0$, common curve for $\tilde{\alpha}=0.05$, Thick curve for $\tilde{\alpha}=0.1$} \label{fg3}
\end{figure*}

Now the ratio $\frac{P_cr_c}{T_c}$ can be calculated as
\begin{equation}
\left.\frac{P_cr_c}{T_c}\right|_{k=1,Q=1}=\frac{21-48\tilde{\alpha}-\sqrt{9-96\tilde{\alpha}+192\tilde{\alpha}^2}}{96(1-2\tilde{\alpha})}
.\label{29}
\end{equation}
When $\tilde{\alpha}=0$, Eq. (\ref{29}) reduces to $\frac{3}{16}$, recovering the result of RN-AdS black holes~\cite{Kubiznak}. When $\tilde{\alpha}\neq0$, the ratio is no longer a constant but a function of $\tilde{\alpha}$, showing the effect of conformal anomaly.

The case $k=-1$ can be discussed similarly. The positive Hawking temperature means
\begin{equation}
\left.T_c\right|_{k=-1,Q=1}=\frac{-r_c^2+4\tilde{\alpha}-2}{2\pi r_c(r_c^2+12\tilde{\alpha})}>0
,\label{30}
\end{equation}
which can be solved as
\begin{equation}
0<r_c<\sqrt{4\tilde{\alpha}-2}.\label{31}
\end{equation}
And the curve of $r_c$ vs. $\tilde{\alpha}$ is depicted in Fig.\ref{4}, which shows that no critical points exists under the constraint of the positive Hawking temperature.
\begin{figure*}
\centerline{\subfigure[]{\label{4a}
\includegraphics[width=8cm,height=6cm]{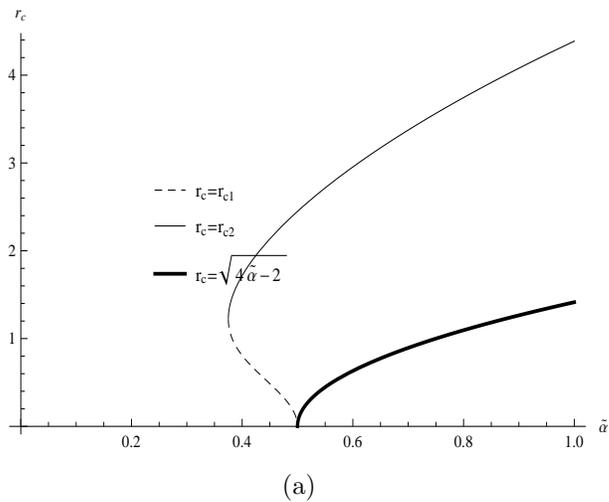}}}
 \caption{$r_c$ vs. $\tilde{\alpha}$ for $k=-1, Q=1$} \label{fg4}
\end{figure*}
\section{Conclusions}
\label{Sec5}
    To observe the effects of conformal anomaly on the thermodynamics of black holes, we focus on the $P-V$ criticality of AdS black holes with conformal anomaly. Specifically, we probe whether conformal anomaly influences the existence of Van der Waals like critical behavior and the critical physical quantities. Treating the cosmological constant as thermodynamic pressure, we extend the former research to the extended phase space. The entropy gains an extra logarithmic term due to the effect of conformal anomaly while the thermodynamic volume is the same as that of RN-AdS black holes.

    Firstly, we study the $P$-$V$ criticality of the uncharged AdS black holes with conformal anomaly. There are two positive roots of $r_c$. However, the Hawking temperature at these two critical points is negative. So these two critical points do not make any sense physically and there would be no Van der Waals like critical behavior for the uncharged case. This is in accord with Schwarzschild AdS black holes, implying that the conformal anomaly does not influence whether there exists Van der Waals like critical behavior. Secondly, we investigate the $P$-$V$ criticality of the charged cases and find that conformal anomaly influences not only the critical quantities but also the ratio $\frac{P_cr_c}{T_c}$.  When $k=1$, with the increasing of conformal anomaly parameter $\tilde{\alpha}$, both $T_c$ and $P_c$ increase while $r_c$ decreases. The ratio $\frac{P_cr_c}{T_c}$ is no longer a constant as before but a function of $\tilde{\alpha}$. The above results show the effects of conformal anomaly. When $k=-1$, there exists no critical points that make any sense physically.

In the context of AdS black holes, $P$-$V$ criticality suggests the existence of first order small/large black hole phase transition below the critical temperature and second order phase transition at the critical point. It resembles the liquid/gas phase transition. By studying the case of nonzero alpha, we show more characteristics other than the common characteristics share by AdS black holes. Not only the critical physical quantities but also the ratio $\frac{P_cr_c}{T_c}$ are influenced due to the effect of conformal anomaly, suggesting that conformal anomaly may affect the small/large black hole phase transition. On the other hand, the curves show that the critical volume/radius exists only for some range of alpha. In this Letter, we have derived the explicit expression of this range. When the conformal anomaly parameter is above this range, one can not find root of critical point that has physical meaning. The physics is not difficult to explain. When $\tilde{\alpha}\rightarrow 0$, the black hole solutions reduces to the RN-AdS black hole. So when the conformal anomaly parameter is small enough, one may expect its critical behavior is not affected too much and similar to RN-AdS black hole. However, when the conformal anomaly parameter is above certain range, the effect of conformal anomaly begins to play an important role, leading to different behavior from RN-AdS black hole.

\ack
 We would like to express our sincere gratitude to both the referees and editors for their hard work and enlightening comments. This work is supported by the National Natural Science
Foundation of China (Grant Nos.11235003, 11175019, 11178007). It is
also supported by \textquotedblleft Thousand Hundred
Ten\textquotedblright \,Project of Guangdong Province and supported by Department of Education of Guangdong Province (Grant No.2014KQNCX191).

\end{document}